# Low temperature Raman study of $Ca_3Co_2O_6$

Vinod panchal , Smita Gohil and Shankar Ghosh

Department of Condensed Matter Physics and Materials Science
Tata Institute of Fundamental Research
Homi Bhabha Road, Mumbai 400-005, India

Abstract

We present low temperature Raman measurements on Ca3Co2O6. Shell model lattice dynamics calculations were done to determine the vibrational pattern of the expected Raman modes. In the temperature window 300K to 52K the mode frequencies and the full width at half maxima (FWHM) of the modes vary in an expected fashion, i.e., the mode frequencies shift to higher values and the FWHM decreases with lowering of temperature. At about 52K ($T^*$) there happens an abrupt change in both the peak position and FWHM. The peak positions show a sudden softening while FWHM increases abruptly. To explain these spectral features we postulate that below $T^*$ the degeneracy of the high symmetry Eg is lifted due to the possible distortion of the $Co^{O}$O6 octahedra.

## Introduction

In Ca3Co2O6 the Co2O6 chains run parallel to the ‘c’ axis with the Ca ions being placed between them. The connectivity between the chains is extremely weak and this gives the chains its one dimensional nature. This shows up in the materials unusual magnetic properties [1–4] and have recently attracted lot of attention [5, 6]. The magnetic exchange in this compound is ferromagnetic (FM) along the chains and antiferromagnetic in the ‘ab’ plane. The specific heat [7] and magnetization investigations [4] in Ca3Co2O6 have suggested that at about 25 K inter-chain antiferromagnetic ordering takes place in the ab-plane. This leads to geometrical frustration of the magnetic structure. Based on increase in the M¨ossbauer line widths below 80K in Eu doped Ca3Co2O6 it was recently shown that the low temperature magnetic ordering has a precursor signature to it [8]. Recent extended X-ray absorption fine structure (EXAFS) experiments on Ca3Co2O6 have also shown an increase in the distribution in the Co − O bond length in the CoO6 octahedra below 75K [9]. In this paper we present temperature dependent micro-Raman studies of Ca3Co2O6 to study the role of phonons above and below the magnetic ordering temperature of 25K. We have also done shell model lattice dynamics calculations to determine the vibrational pattern of the expected Raman modes.

Ca3Co2O6 has rhombohedral structure with space group R3c and two formula units per unit cell. The cell dimensions are $a_R$ = 628.1pm and $\alpha$ = 92.568$^{\circ}$ [1,10]. The fractional coordinates are given in table 1. In the rhombohedral settings of R3c the Co ions occupy the 2a(0.25, 0.25, 0.25) and the 2b(0, 0, 0) sites, the Ca ions occupy the 6e(x, −x+0.5, 0.25) sites, while the O ions are in the 12f(x, y, z) sites. The Co2O6 chains running along the c axis of Ca3Co2O6 are are made of two alternately occuring polyhedron (i) octahedra and (ii) trigonal prisms. These two polyhedra share their triangular faces. The Co ions in the 2a site ($Co^{O}$) participate in the octahedral coordination with the oxygen atoms while the Co ions in the 2b site ($Co^{Tr}$) are in the trigonal prismatic site. The $Co^{O}$ − O(∼ 191.6pm) distance is smaller than $Co^{Tr}$ − O(∼ 206.2pm) indicating presence of charge ordering along the chains. Both the spin and valence state of the $Co^{Tr}$ and $Co^{O}$ have been a matter of debate. It was initially suggested that $Co^{O}$ has a charge of $4^{+}$ while the $Co^{Tr}$ has a charge of $2^{+}$ [1], however recent x-ray absorption and magnetic circular dichroism suggests that both $Co^{O}$ and $Co^{Tr}$ ions have the the same charge of $3^{+}$ [11]. On similar lines it was also shown that the $Co^{O}$ is in the low spin state while the $Co^{Tr}$ is in the high spin state [12].

Of the 66 zone-center phonon modes 3 are acoustic (A2u + Eu) 24 are Raman active (4A1g+10Eg), 29 are IR active (6A2u+11Eu) and 11 are silent (5A1u+6A2g) [13]. All the atoms participate in the IR active modes while in the Raman active A1g mode none of the Co ions move and in the other Raman active E1g the $Co^{O}$ site is stationary [13].

Polycrystalline samples of Ca3Co2O6 were prepared by a solid state method starting from stoichiometric amounts of high purity (> 99.99%) CaCO3 and Co3O4 powder. The specimen obtained after few heat treatments were found to be single phase by x-ray (Cu K ) diffraction measurements. Raman scattering measurements were performed in the back scattering geometry using a Jobin Yvon T64000 Raman spectrometer equipped with an inverted microscope. Raman spectra were recorded using 514.5 nm radiation of an argon-ion laser with a power of 5 mW. The temperature range of 300 to 4.2 K was covered using a continuous-flow helium cryostat from M/s Oxford, UK.

The ambient condition (300K) Raman spectrum is shown in Fig.1. It can be seen that the observed Raman spectra (shown by filled circles) can be fitted to a sum of five-Lorentzians. The individual components are shown by dotted lines. The five modes are : 312 $cm^{-1}$[W1], 456 $cm^{-1}$[W2 ]), 529$cm^{-1}$[W3 ], 562 $cm^{-1}$[W4] and 629$cm^{-1}$[W5 ] . The polycrystalline samples used consisted of irregularly shaped optically isotropic grains and hence polarization dependent mode assignment was not possible in the sample. In absence of single crystals the mode assignments were guided by lattice dynamics calculations. Before we present the temperature dependence of the modes, we will present the lattice dynamic calculations to assign the observed Raman bands.

The lattice-dynamics calculations for the phonons of Ca3Co2O6 at the Brillouin zone center were carried out using shell-model [14] . The long-range interactions between each pair of ions was taken to be of Coulombic form, while the short range interactions between the shells was taken to be of Born-Mayer-Buckinghan form: $V_{kk'} (r_{ij})= A_{kk'} \exp (-r_{ij}/R_{kk'} ) - C_{kk'} /r^{6}$, where $r_{ij}$ is the distance between the two ions i and j of species k and k′, respectively. The parameters Akk′ ,$R_{kk'}$ and $C_{kk'}$ are given in Table 2. In the shell model, each ion of charge Z|e| is represented by a massless shell of charge Q|e| and a core of charge Q =(Z −Y )|e|. The shells are elastically bound to the cores with a force constant K whose values are given in Table 1. it can be seen that to get

reasonable agreement to the experimental data we had to treat $Co^{O}$ and $Co^{Tr}$ separately giving an ionic charge 3.135 (2.56) to $Co^{O}$($Co^{Tr}$). Based on eigenvector analysis the fourteen Raman active modes can be broadly classified as (a) rotational (b) bending (c) breathing and (d)anti-stretching of the $Co^{O}$O6 octahedra. Schematically this is shown in Fig.2. These classifications are consistent with the one performed for rhombohedral LaAlO3 and LaMnO3 [15]. Based on the lattice dynamics calculations and also guided by the mode analysis of rhombohedral LaAlO3 and LaMnO3 [15] we associate the modes W1 and W2 to the 'anti-stretching' W3 to 'rotational' and W5 to the 'bending' mode. The W4 mode remains unidentified.

We now present the low temperature results. Inside the cryostat only few Raman modes were clearly discernible at all temperatures. Hence only these modes .i.e., W3,W4,W5 were analyzed. In the temperature range 300K to 52 K the temperature dependence of the mode frequencies and their full width at half maxima (FWHM) change in an expected fashion, i.e., the mode frequencies shift to higher values and the FWHM decreases with lowering of temperature. At about 52K ($T^{*}$) there happens an abrupt change in both the peak position and FWHM. To highlight this change at 52K all the data has been plotted in the temperature range 100Kto 5K. Figure 3 shows representative Raman spectra at few select temperatures. It can be easily seen that the that that the Raman modes do not monotonically shift to higher frequency with lowering of temperature, at about 52K ($T^{*}$) the modes suddenly soften and this is also associated with the sudden broadening of the peaks. The dashed straight lines in Fig.3 which are a guide to the eye highlight this feature. Figure 4 and 5 show the temperature dependence of the peak positions and the FWHM of the W3,W4,W5 Raman modes respectively . It can be seen that at $T^{*}$(~ 52K) (marked by the dashed line) there is a pronounced drop in peak positions. The dramatic change in the peak position of the modes is also accompanied by a simultaneous broadening of their corresponding FWHM (Notice that all the modes at 52K are very broad). The solid lines in the figures are linear fits to the data.

In conclusion our experiments show that at about 52K ($T^{*}$) the Raman modes (characterized to be the bending modes of the $Co^{O}$O6 octahedra) the modes suddenly soften and this is also associated with the sudden broadening of the peaks. This we associate with the appearance of new modes caused by the distortion of the $Co^{O}O6$ octahedra which sets up a a freezing of a breathing phonon mode type distortion. Comparing our results with that obtained from CaFeO3we speculate that such lattice distortion could initiate a charge redistribution on the Co2O6 chains. Whether such a reorganization of charge can set up a precursor signature to a magnetic ordering is not obvious. These inferences based on the Raman data is also consistent with the EXAFS data which shows increase in the Co − O bond length distribution below 75K. Interestingly there exists an inconsistency in the $T^{*}$ from the Raman and the EXAFS measurements [9]. While the $T^{*}$ for the EXAFS is about 75K [9] the one obtained from Raman is 52K. This inconsistency in $T^{*}$ can not be attributed to laser heating of the sample alone. In the semi empirical lattice dynamics calculations the $Co^{O}$ and $Co^{Tr}$ atoms were needed to be treated separately, this not consistent with the x-ray absorption and magnetic circular dichroism suggests that both $Co^{O}$ and $Co^{Tr}$ ions have the the same charge of $3^{+}$ [11]. One needs possibly an ab initio lattice dynamics calculations to resolve the issue.

The authors thank and Prof. E. V. Sampathkumaran for useful discussions. SG thanks Prof. K. Maiti for helpful discussions and sharing his unpublished data of EXAFS on Ca3Co2O6.

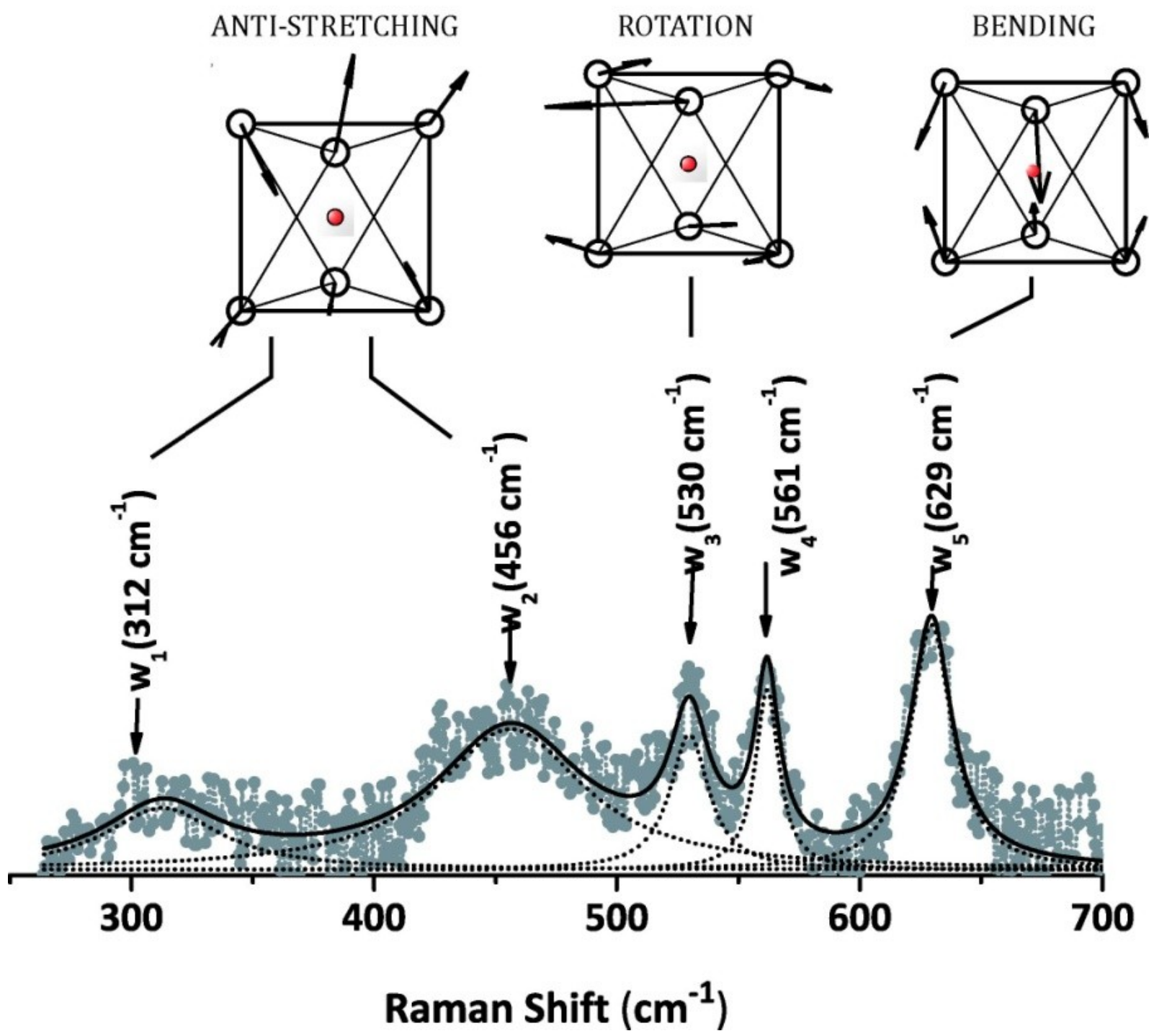

Fig. 1: Room temperature Raman spectra of Ca3Co2O6 . The ambient spectra shows five modes (W1 ... W5). The dotted lines are Lorentzian fits to the data. The W1 has been classified as rotational mode while W2 ... W5 has been classified as bending mode of the $Co^{O}$O6 octahedra. The modes are schematically shown in the figure, we associate the modes W1 and W2 to the 'anti-stretching' W3 to 'rotational' and W5 to the 'bending' mode. The W4 mode remains unidentified.

| ion | Wyckoff Notation | Site Symmetry | x | y | z | Irreducible representantions |
|---|---|---|---|---|---|---|
| Ca | 6e | C2 | 0.6205 | 0.8795 | 0.25 | $A_{1g} + A_{1u} + 2A_{2g} + 2A_{2u} + 3E_u + 3E_g$ |
| $Co^{O}$ | 2a | D3 | 0.25 | 0.25 | 0.25 | $A_{2g} + A_{2u} + E_u$ + Eg |
| $Co^{Tr}$ | 2b | S6 | 0 | 0 | 0 | $A1_u + A_{2u} + E_u$ |
| O | 12e | C1 | 0.2912 | 0.9605 | 0.0909 | $3A_{1g} + 3A_{1u} + 3A_{2g} + 3A_{2u} + 6E_u$ |

Mode Classification

| | |
|---|---|
| ΓRaman = 4A1g + 10Eg | ΓIR = 6A2u + 11Eu |
| ΓAcoustic = A2u + Eu | ΓSilent = 5A1u + 6A2g |

Table 1: Wyckoff notations, site symmetries, fractional atomic coordinates and irreducible representation for the atoms

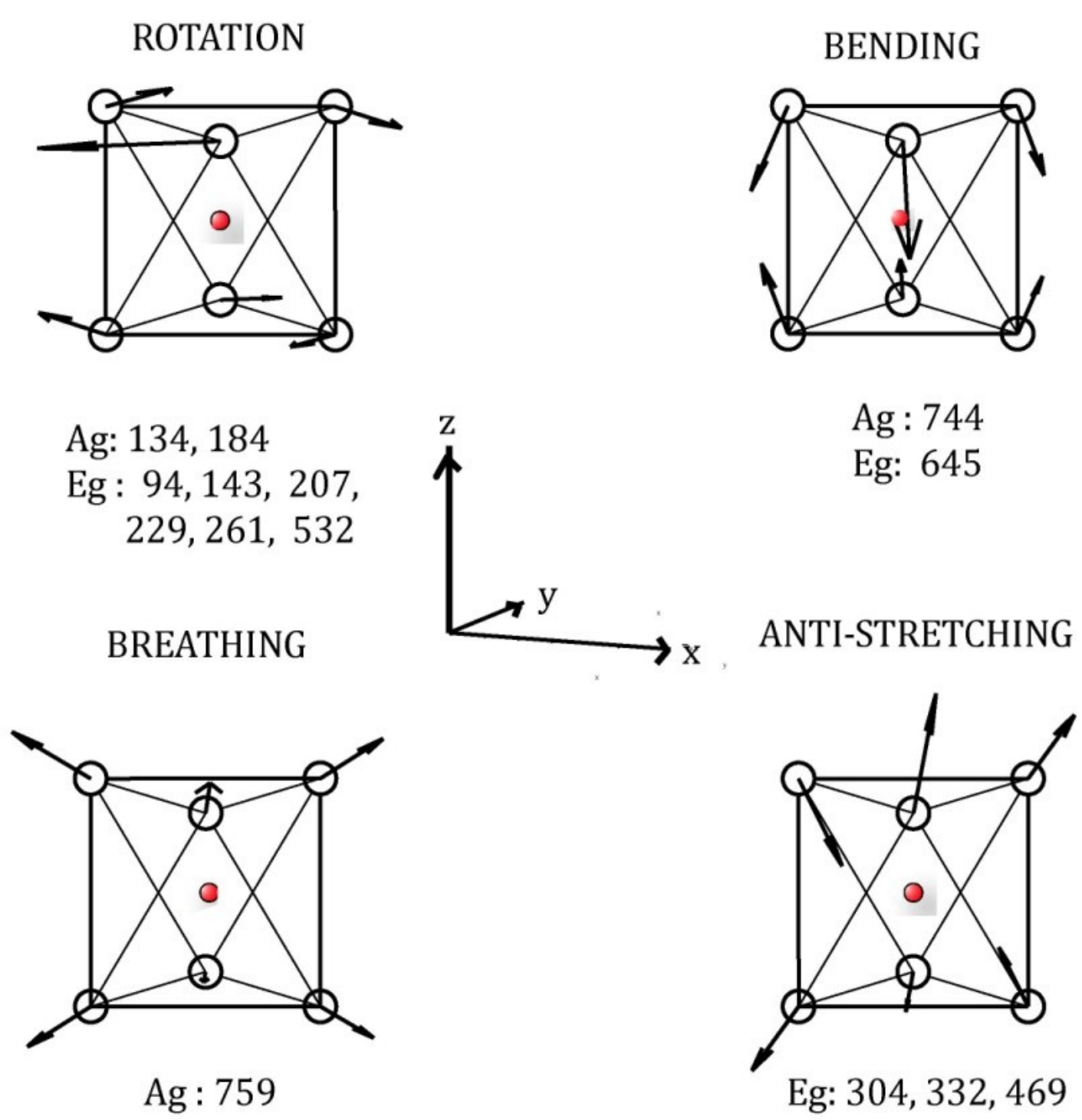

Fig. 2: The calculated vibrational patterns and frequencies of the Raman active optical phonons in Ca3Co2O6. For the sake of clarity only the $Co^{O}$ O6 octahedron is drawn . In the Ag mode none of the the Co ions move, however, in the Eg mode the $Co^{Tr}$ (not shown ) moves.

| ion | Y (core) e | Q (shell) e | K(k) N/m | ionic pair | $A_{kk'}$ eV | $R_{kk'}$ ˚A | $C_{kk'}$ ion eV ˚$A_6$ |
|---|---|---|---|---|---|---|---|
| Ca | -1.6 | 3.5 | 1882 | O − Ca | 1227 | 0.34 | 0 |
| $Co_O$ | -0.665 | 3.8 | 121910 | O − Co | 790 | 0.34 | 0 |
| $Co_{Tr}$ | -1.2350 | 3.8 | 121910 | O − $Co_{Tr}$ | 700 | 0.34 | 0 |
| O | 1.1 | -3 | 1037 | O − O | 22764 | 0.15 | 27.87 |

Table 2: Parameters of the shell model described in the text. Y(k), the core Q(k), shell charge; K(k), core-shell force constant. $A_{kk'}$ ,$R_{kk'}$ and $C_{kk'}$ parameters of the Born-Mayer-Buckinghan potential.

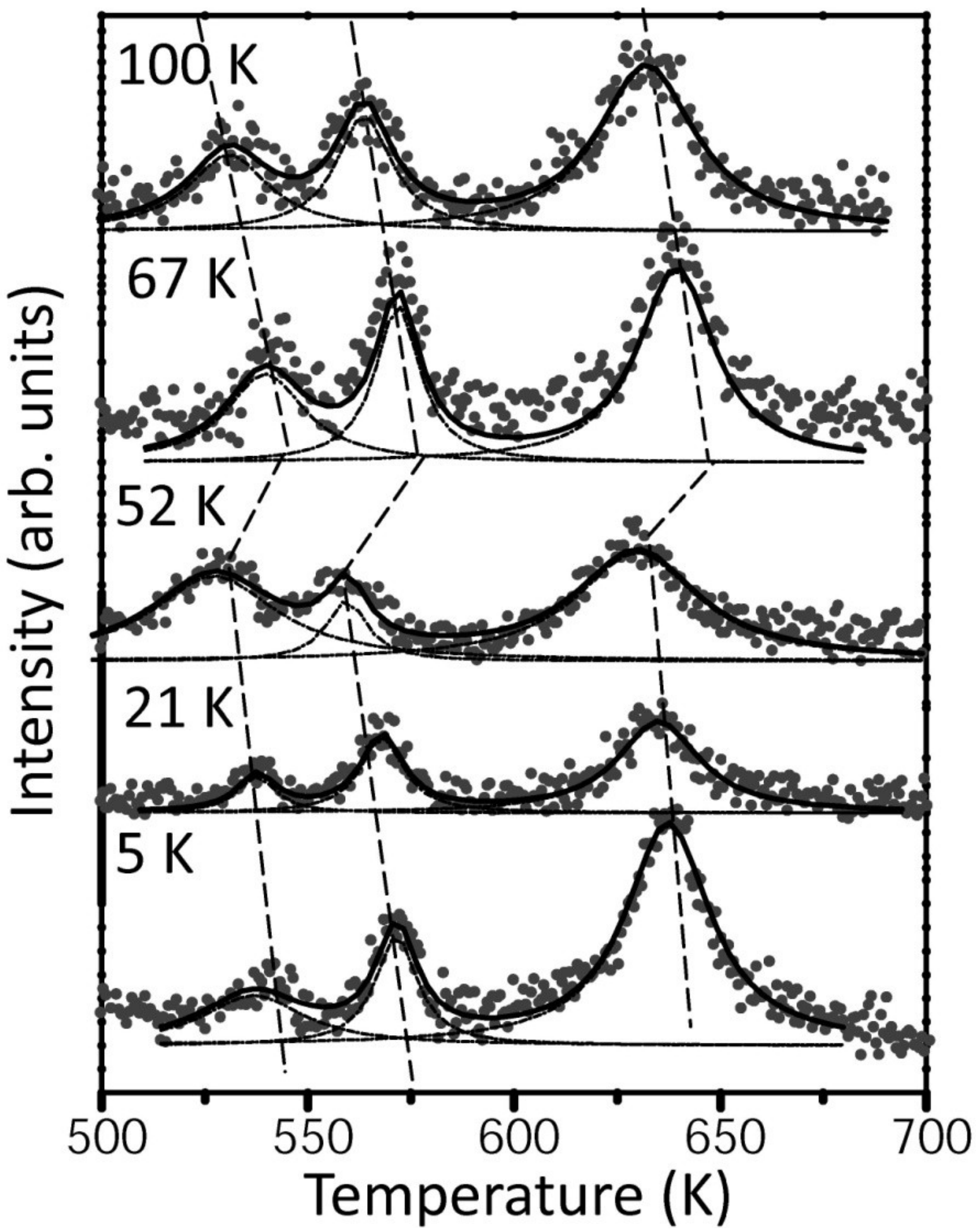

Fig. 3: Raman spectra of Ca3Co2O6 for a few selected temperatures ( shown by the side of the spectrum) in the temperature range 100 to 4 K. The dotted lines show Lorentzian fits to the data.

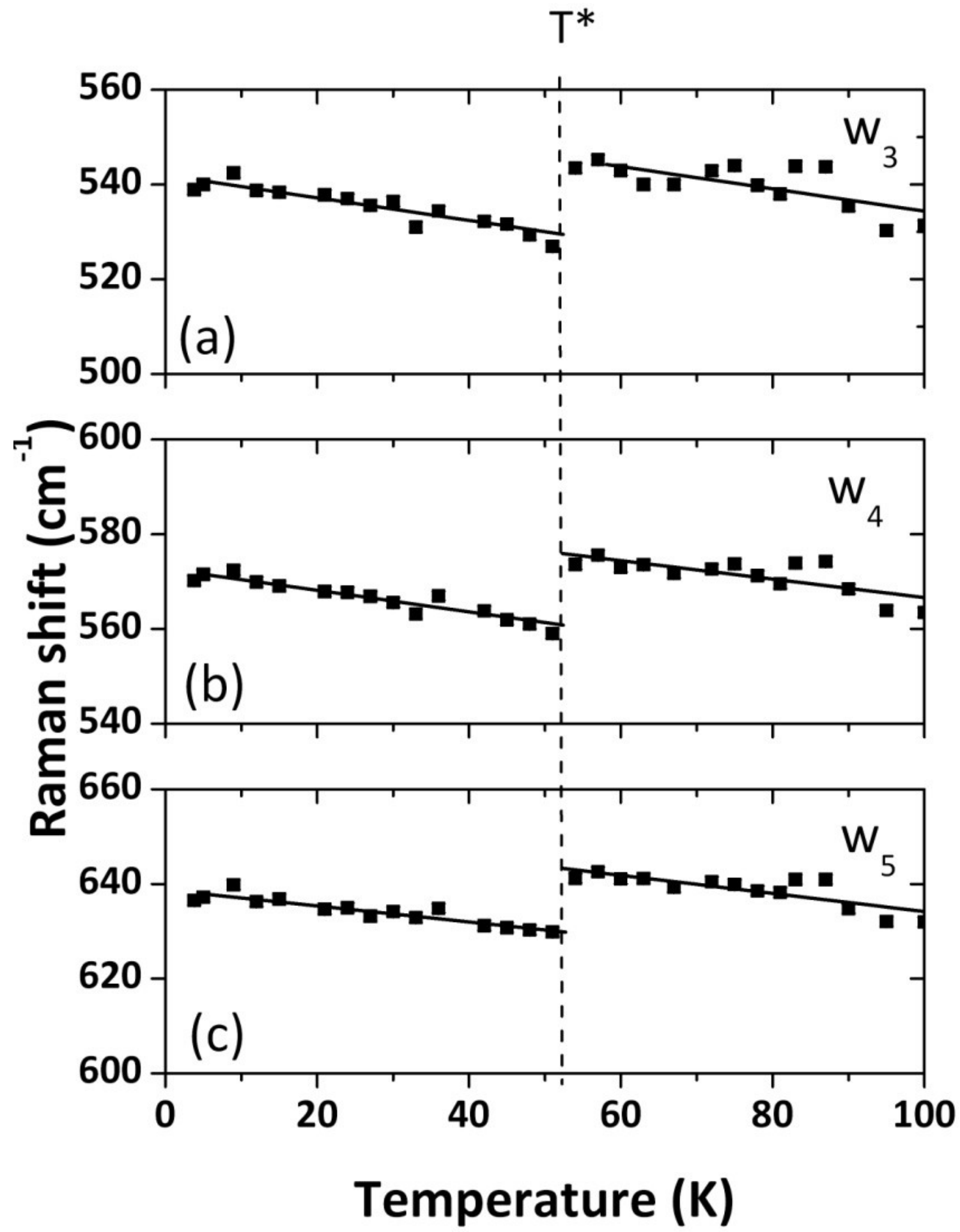

Fig. 4: Temperature dependence of the peak positions of the modes W3,W4,W5. It can be seen that at $T^*$ (~ 52K) (marked by the dashed line) there is a sudden drop in peak positions. The solid line is a linear fit to the data. The slope of the solid line for all the modes is about -0.2 $cm^{-1}$/K both above and below $T^*$

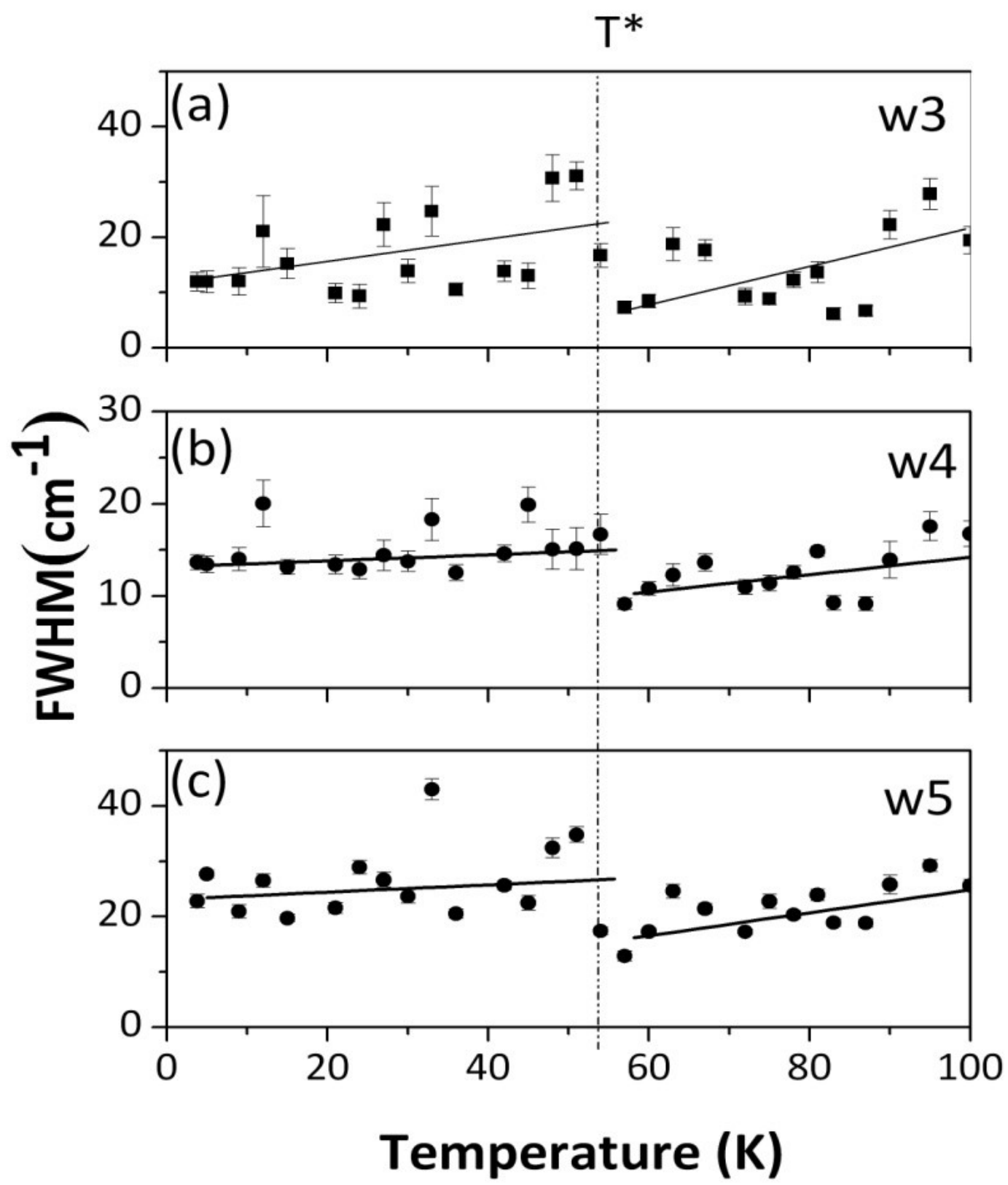

Fig. 5: Temperature dependence of the FWHM of the modes W3,W4,W5. It can be seen that at about 52 K (marked by the dashed line) there is a sudden increase in FWHM. The solid lines are a linear fit to the data.